\documentclass[conference]{IEEEtran}
\IEEEoverridecommandlockouts
\usepackage{cite}
\usepackage{amsmath,amssymb,amsfonts}
\usepackage{algorithmic}
\usepackage{graphicx}
\usepackage{textcomp}
\usepackage{xcolor}
\usepackage[pscoord]{eso-pic}
\newcommand{\placetextbox}[3]{
  \setbox0=\hbox{#3}
  \AddToShipoutPictureFG*{
    \put(\LenToUnit{#1\paperwidth},\LenToUnit{#2\paperheight}){\vtop{{\null}\makebox[0pt][c]{#3}}}%
  }%
}%
\def\BibTeX{{\rm B\kern-.05em{\sc i\kern-.025em b}\kern-.08em
    T\kern-.1667em\lower.7ex\hbox{E}\kern-.125emX}}
\def\BibTeX{{\rm B\kern-.05em{\sc i\kern-.025em b}\kern-.08em
    T\kern-.1667em\lower.7ex\hbox{E}\kern-.125emX}}

\usepackage{csquotes}
\usepackage{textgreek}
\usepackage{tabularx}
\usepackage{multirow}
\usepackage{multicol}
\usepackage{soul}
\usepackage{graphicx} 
\usepackage{caption}

\usepackage{etoolbox}
\def\BibTeX{{\rm B\kern-.05em{\sc i\kern-.025em b}\kern-.08em
    T\kern-.1667em\lower.7ex\hbox{E}\kern-.125emX}}

\makeatletter
\patchcmd{\@maketitle}
  {\addvspace{0.5\baselineskip}\egroup}
  {\addvspace{-1.2\baselineskip}\egroup}
  {}
  {}

\usepackage{dblfloatfix}    
\usepackage{textcomp}

\usepackage{booktabs} 
\usepackage{graphicx}
\graphicspath{{figs/}}
\usepackage{braket}
\usepackage{multirow}

\usepackage{cite}
\usepackage{amsmath,amssymb,amsfonts}
\usepackage{algorithmic}
\usepackage{graphicx}
\usepackage{textcomp}
\usepackage{color}
\def\BibTeX{{\rm B\kern-.05em{\sc i\kern-.025em b}\kern-.08em
    T\kern-.1667em\lower.7ex\hbox{E}\kern-.125emX}}

\usepackage{fancyhdr}

\usepackage{lipsum}
\usepackage{marginnote}

\begin{document}

\placetextbox{0.5}{1}{Accepted in the Design Automation and Test in Europe Conference 2020}%

\title{Accelerating Quantum Approximate Optimization Algorithm using Machine Learning
}

\author{\IEEEauthorblockN{Mahabubul Alam}
\IEEEauthorblockA{\textit{Department of Electrical Engineering} \\
\textit{Pennsylvania State University}\\
University Park, USA \\
mxa890@psu.edu}
\and
\IEEEauthorblockN{Abdullah Ash-Saki}
\IEEEauthorblockA{\textit{Department of Electrical Engineering} \\
\textit{Pennsylvania State University}\\
University Park, USA \\
ash.saki@psu.edu}
\and
\IEEEauthorblockN{Swaroop Ghosh}
\IEEEauthorblockA{\textit{Department of Electrical Engineering} \\
\textit{Pennsylvania State University}\\
University Park, USA \\
szg212@psu.edu}
}

\maketitle

\begin{abstract}
We propose a machine learning based approach to accelerate quantum approximate optimization algorithm (QAOA) implementation which 
is a promising quantum-classical hybrid algorithm to prove the so-called \emph{quantum supremacy}. In QAOA, a parametric quantum circuit and a classical optimizer iterates in a closed loop to solve hard combinatorial optimization problems. 
The performance of QAOA improves with increasing number of stages (depth) in the quantum circuit. However, two new parameters are introduced with each added stage for the classical optimizer increasing the number of optimization loop iterations.
We note a correlation among parameters of the lower-depth and the higher-depth QAOA implementations and, exploit it by developing a machine learning model to predict the gate parameters close to the optimal values. As a result, the optimization loop converges in a fewer number of iterations. We choose graph MaxCut problem as a prototype to solve using QAOA. We perform a feature extraction routine using $100$ different QAOA instances and develop a training data-set with $13,860$ optimal parameters. We present our analysis for $4$ flavors of regression models and $4$ flavors of classical optimizers. Finally, we show that the proposed approach can curtail the number of optimization iterations by on average $44.9\%$ (up to $65.7\%$) from an analysis performed with $264$ flavors of graphs.

\end{abstract}


\section{Introduction}


Quantum approximate optimization algorithm (QAOA) \cite{farhi2014quantum, farhi2017quantum} is a quantum-classical hybrid algorithm to solve combinatorial optimization problems. 
The hybrid algorithms are considered promising to demonstrate quantum advantage (i.e., to prove superior performance for a problem compared to state-of-the-art classical methods) \cite{farhi2016quantum, preskill2018quantum}. 
Versions of the QAOA are expected to find approximate solutions to combinatorial search problems faster than the classical algorithms. 
Thus, exploring new avenues to improve the performance of QAOA has become an exciting research domain \cite{zhou2018quantum, crooks2018performance, wilson2019optimizing, wecker2016training, streif2019training}.

\begin{figure*} [!ht] 
\vspace{-1em}
 \begin{center}
    \includegraphics[width=0.95\textwidth]{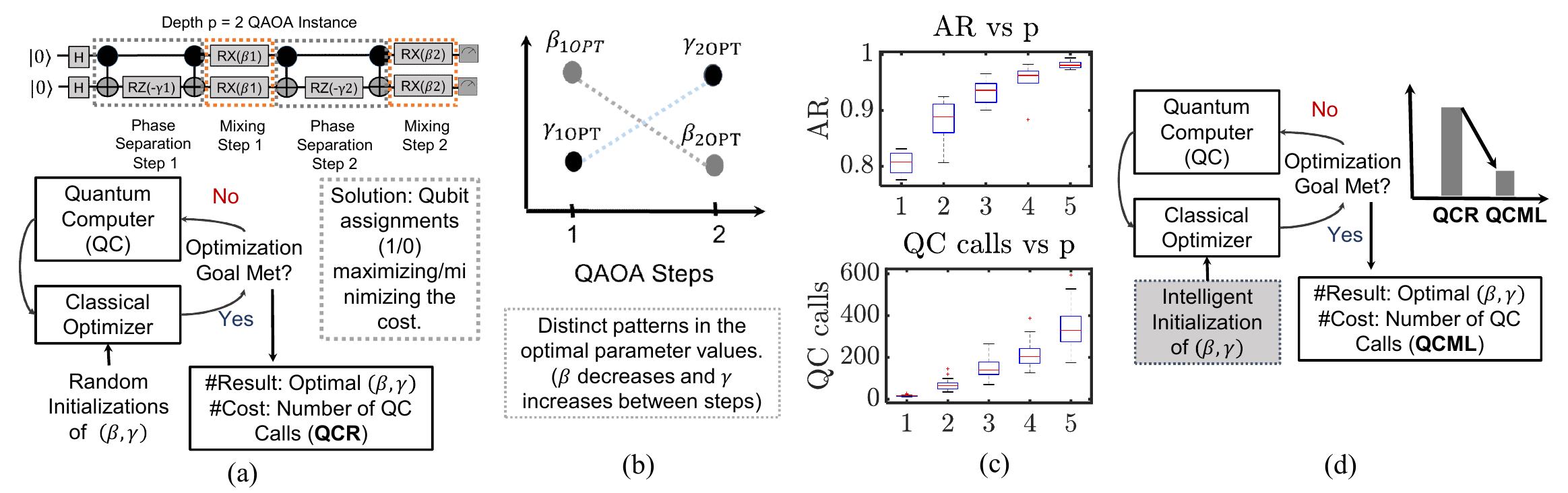}
 \end{center}
 \vspace{-1em}
 \caption{(a) Typical QAOA flow with random initializations (QCR) of the control parameters and a QAOA circuit instance for a MaxCut problem of two-node, single edge graph; (b) expected patterns in the optimal parameters; (c) approximation ratio - AR (indicates the performance) and run-time (QC calls) distributions for QAOA MaxCut instance optimization of four 3-Regular graphs (8-nodes) with varying depths ($p$); (d) proposed QAOA flow with ML-based initialization (QCML) of the parameters.} \label{fig:illustration}
 \vspace{-0.5em}
\end{figure*}

The theoretic discussion of QAOA can be found in \cite{farhi2014quantum, zhou2018quantum}. In this paper, we delineate our proposal based on the quantum circuit-level implementation of the QAOA. Fig. \ref{fig:illustration}(a) shows a typical stage of the QAOA circuit. The first layer consists of Hadamard gate which put the qubits in superposition states. The next level is the phase-separation layer consisting of Controlled-NOT (CNOT) and parametric $RZ(-\gamma)$ gates. 
The final layer is the mixing layer with parametric $RX(\beta)$ gates. Thus, each layer of the circuit has two gate parameters $(\gamma, \beta)$. 
This typical stage can be repeated several times to construct a higher-depth QAOA circuit where each stage will have different sets of gate parameters ($\gamma$ and $\beta$). The QAOA performance is known to improve with the increasing number of stages in the circuit \cite{zhou2018quantum, crooks2018performance}. On the basis of the gate parameters, the quantum circuit generates an output quantum state $\ket{\psi({\gamma}, {\beta})}$. QAOA involves a cost function which is the expectation value of cost Hamiltonian ($H_C$) in the output state $\ket{\psi(\gamma, \beta)}$. The optimization goal is to maximize this expectation value. 
In each iteration the classical optimizer tracks the expectation value and generates a new set of parameters $(\gamma, \beta)$ which drives the quantum circuit.

The optimization loop iterates between the quantum computer and the classical optimizer until an optimal set of gate parameters are found (Fig. \ref{fig:illustration}(a)). The number of this loop-iteration is a key factor for the QAOA run-time. A higher-depth (stage) QAOA implementation has better performance but a larger number of parameters may lead to a greater number of loop iterations (Fig. \ref{fig:illustration}(c)). (\# of loop iterations, function calls and QC calls are used interchangeably throughout the paper.)

In a straight-forward method, the optimization loop starts from a random set of gate parameters (noted as QCR in Fig. \ref{fig:illustration}(a)) which requires higher number of optimization loops. 
We observe correlations among the gate parameters (as depicted in Fig. \ref{fig:illustration}(b)), analyze these trends, extract \emph{features}, and train a Machine Learning (ML) based predictor model. The predictor model predicts initial parameters for a higher-depth QAOA circuit from the optimal gate parameters of a single-stage QAOA. These tuned initial parameters are close to the optimal ones. Therefore, when the optimization loop is \emph{intelligently initialized} (noted as QCML in Fig. \ref{fig:illustration}(d)) with these tuned parameters, the optimization goal is achieved faster cutting down the loop iterations by 44.9\% on average (up to 65.7\%).


We select MaxCut problem to be solved by QAOA. The backgrounds of MaxCut formulation and QAOA are provided in the Appendix. The contributions made in this paper are:

\noindent \textbf{(a) Feature extraction:} We explore a graph MaxCut under various setups (e.g., 100 different QAOA instances with varied depth) and identify optimal gate parameters for each setup. From this optimal values we select features for our ML model. 

\noindent \textbf{(b) Training data-set:} We prepare a training data-set with a total of $13,860$ optimal parameters ($330$ different graphs and $6$ QAOA instances for each graph).

\noindent \textbf{(c) ML model:} We train $4$ ML models, namely, Gaussian Process Regression (GPR), Linear Regression (LM), Regression Tree (RTREE), and Support Vector Machine Regression (RSVM) to study the effect of ML model on the classical loop optimization. GPR exhibits best performance on the basis of prediction accuracy.  

\noindent \textbf{(d) Accelerating optimization loop:} We propose a two-level ML-based approach to accelerate QAOA optimization loop. 
First, we calculate the optimal parameters $(\gamma_{1OPT}, \beta_{1OPT})$ for a single-stage (lowest depth) QAOA circuit using naive method. Next, we feed this $(\gamma_{1OPT}, \beta_{1OPT})$ of the single-stage implementation to ML model and predict tuned parameters for a higher target depth QAOA instance. The higher depth instance (initialized with the predicted parameters) is then run in the optimization loop with a local optimizer to generate the final solution.
We also introduce a hierarchical prediction by using optimal parameters from an intermediate-stage QAOA implementation along with the single-stage values. 

\noindent \textbf{(e) Exhaustive analysis:} We explore a broad-spectrum of classical optimizers (a total of $4$) to establish that our approach is optimizer-agnostic. Two of these optimizers are gradient-based (L-BFGS-B and SLSQP), and two are gradient free (Nelder-Mead and COBLYA) from Python SciPy library \cite{jones2016scipy}). We have used QuTIP library \cite{johansson2013qutip} based quantum computer simulator as the quantum computer of the optimization loop.

\noindent \textbf{(f) Quantified speed-up:} The proposed approaches reduce the optimization loop iteration by 44.9\% on average.

\emph{To the best of our knowledge, this is the first work on optimizing the parameters of QAOA using ML}.

\section{Parameter Trends, Feature Selection, and Proposed Approach} 
In this section, we describe the notations used in the paper to avoid ambiguity. We also discuss the trends in the QAOA circuit parameters that are leveraged to select appropriate features for the ML model and to predict the initial parameters. 
\subsection{Notations}
The depth (or, the \emph{total} number of stages) in a QAOA circuit is denoted by $p$. Each stage/step of a $p$-depth circuit is indexed with $i$. For example, for a QAOA circuit with depth $(p=)$ $5$, $i$ could be $\{1,2,3,4,5\}$. For a single-depth circuit (i.e., $p = 1$), $i=\{1\}$. Additionally, $\gamma_{1OPT(p=3)}$ denotes $\gamma$ parameter (i.e., phase separation parameter) of the $1^{st}$ stage $(i=1)$ of a QAOA circuit implementation with depth $3$. Note that, the same problem can be solved by lower depth QAOA circuit as well as by a higher depth QAOA circuit. The Approximation Ratio (AR) of the higher depth implementation is better. 
 
\subsection{Patterns in Optimal Control Parameters} \label{obs1}
\begin{figure} [b] 
\vspace{-1em}
 \begin{center}
    \includegraphics[width=0.45\textwidth]{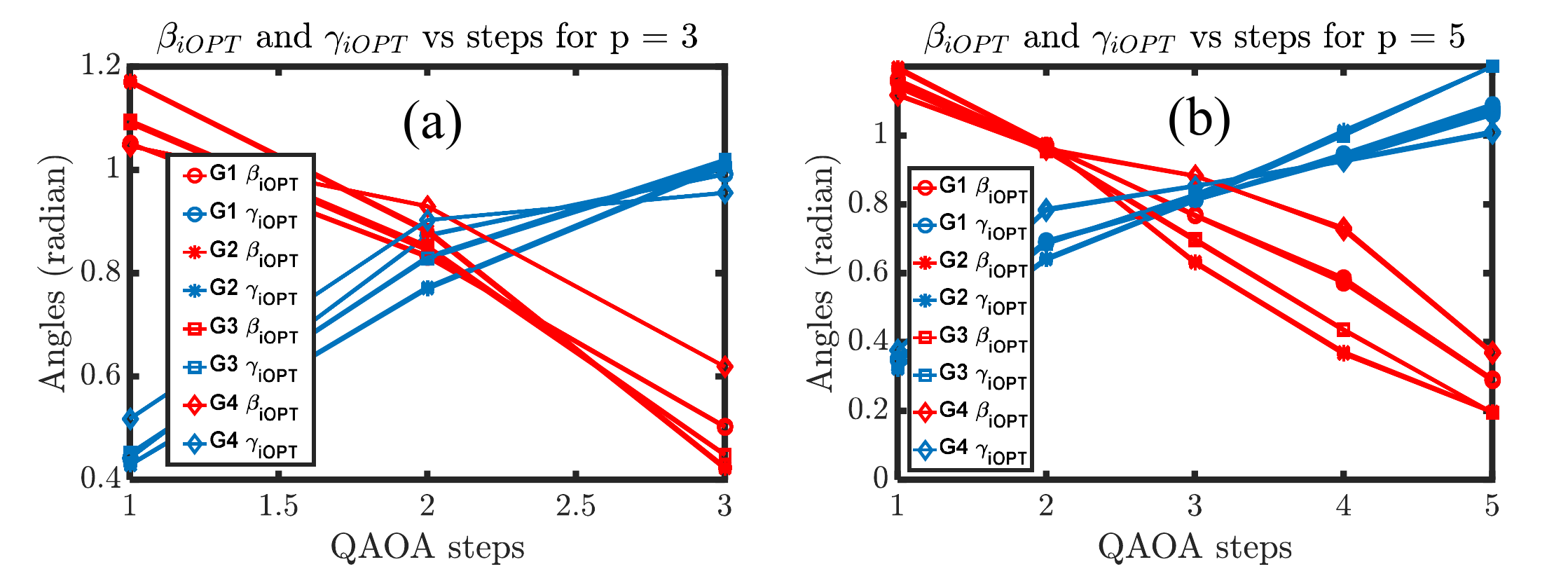}
 \end{center}
 \vspace{-1em}
 \caption{Trends in the optimal control parameters of four 3-regular graphs for (a) $p$ = 3; (b) $p$ = 5 (each instance is optimized from 20 random initializations).} \label{fig:lin}
 \vspace{-0.5em}
\end{figure}

In this Section, we present the patterns in the optimal control parameters for a fixed-depth QAOA circuit (i.e. patterns in the optimal $\gamma_{1OPT}$, $\gamma_{2OPT}$, and $\gamma_{3OPT}$ values for QAOA instance with $p$ = 3). An interesting observation is that the optimal parameter values of any QAOA-instance are not necessarily random. Rather, they show some regularities \cite{zhou2018quantum, crooks2018performance}. The optimal mixing layer parameter ($\beta_{iOPT}$) values of any QAOA instance with depth-$p$ gradually decrease between steps (stages) whereas the optimal phase separating parameters ($\gamma_{iOPT}$) increase between steps. Fig. \ref{fig:lin}(a) and (b) show the optimal control parameter values at 3 steps ($\gamma_{1OPT(p=3)}$, $\beta_{1OPT(p=3)}$ .. $\gamma_{3OPT(p=3)}$, $\beta_{3OPT(p=3)}$) and 5 steps ($\gamma_{1OPT(p=5)}$, $\beta_{1OPT(p=5)}$ .. $\gamma_{5OPT(p=5)}$, $\beta_{5OPT(p=5)}$) for four (denoted by G1, G2, G3 and G4 in Fig. \ref{fig:lin}) 8-node 3-regular graphs with $p$ = 3 and $p$ = 5 respectively. For every graph and every $p$-value, L-BFGS-B optimizer has been used with functional tolerance limit of $10^{-6}$ to find the optimal parameters from 20 random initialization.

\subsection{Optimal Control Parameters with Depth-p} \label{obs2}
\begin{figure} [] 
\vspace{-1em}
 \begin{center}
    \includegraphics[width=0.45\textwidth]{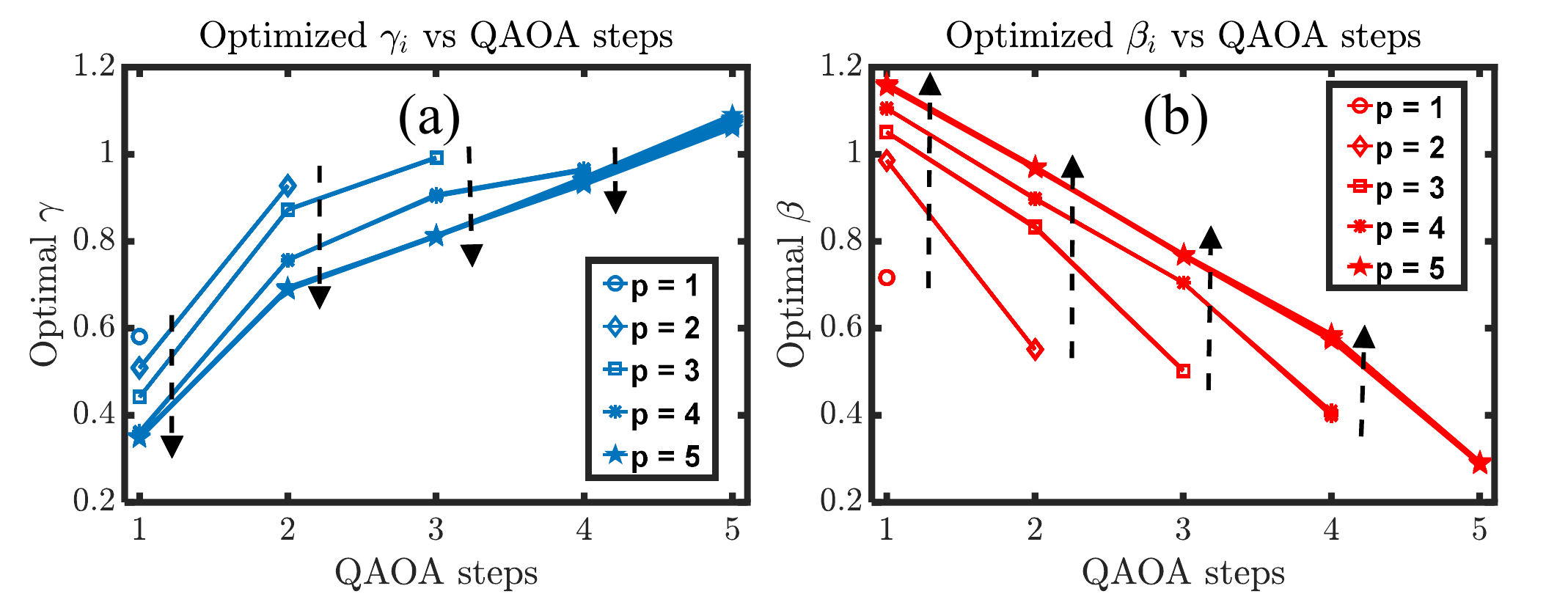}
 \end{center}
 \vspace{-1em}
 \caption{Trends in (a) the optimal $\gamma_{iOPT}$ values, and (b) the optimal $\beta_{iOPT}$ values for varying depths for a single 3-regular graph with 20 random initializations.} \label{fig:single_bg}
 \vspace{-0.5em}
\end{figure}

In this Section, we analyze the patterns in a certain control parameter (say, $\gamma_{1OPT}$) for different instance depths (i.e. how $\gamma_{1OPT}$ value changes from $p$ = 1 ($\gamma_{1OPT}(p=1)$) to $p$ = 2 ($\gamma_{1OPT}(p=2)$) for the same MaxCut problem). The optimal control parameter values of a certain QAOA step ($\gamma_{iOPT}$, $\beta_{iOPT}$) have a strong correlation with the chosen circuit depth-$p$. The optimal phase separating control parameter of a certain QAOA step ($\gamma_{iOPT}$) decreases with the circuit depth-$p$ (annotated by arrows in Fig. \ref{fig:single_bg}(a)). Contrarily, the optimal mixing control parameter ($\beta_{iOPT}$) increases. Fig. \ref{fig:single_bg}(a) shows the optimal phase separating control parameters 
with varying depth ($p$ = 1 to 5). Fig. \ref{fig:single_bg}(b) shows the corresponding optimal mixing parameters.

Note that the above observations are significant. The optimal control parameter values at a lower depth (say, $p1$) gives significant clues about the optimal control parameters at a depth $p2$ where $p2$ $>$ $p1$. For instance, if we have already determined $\gamma_{1OPT}(p=1)$ and $\beta_{1OPT}(p=1)$ for any given problem for $p = 1$, a smaller $\gamma_{1init}(p=2)$ value (than $\gamma_{1OPT}(p=1)$) can be a good starting point for $p=2$ instance optimization (Fig. \ref{fig:single_bg}). Moreover, an initial value larger than $\gamma_{1init}(p=2)$ can be a good starting point for $\gamma_{2init}(p=2)$ (Fig. \ref{fig:lin}). Similar techniques can be applied for the $\beta_{1init}(p=2)$ and $\beta_{2init}(p=2)$ initialization. However, the extent of the differences will depend on the difference between $p2$ and $p1$, and the optimal $\gamma_{1OPT}(p=1)$ and $\beta_{1OPT}(p=1)$ values. A predictor model trained with sufficient number similar problem instances can essentially learn these correlations and can be used for smart initialization of the variables of a higher-depth implementation from the low-depth optimal control parameters.

\subsection{Feature Selection}

The lower-depth optimal control parameters and the target depth (say, $p = p_t$) are used as the $features$ of the predictor models. For the two-level approach (discussed next), $\gamma_{1OPT}(p=1)$, $\beta_{1OPT}(p=1)$, and the target depth $p_t$ are used as the $feature$ vectors (a total of $3$ features). On the basis of these $3$ features, the predictor will generate $2p_t$ parameters. 

\subsection{Proposed Approach to Accelerate QAOA}

\begin{figure} [] 
\vspace{-1em}
 \begin{center}
    \includegraphics[width=0.38\textwidth]{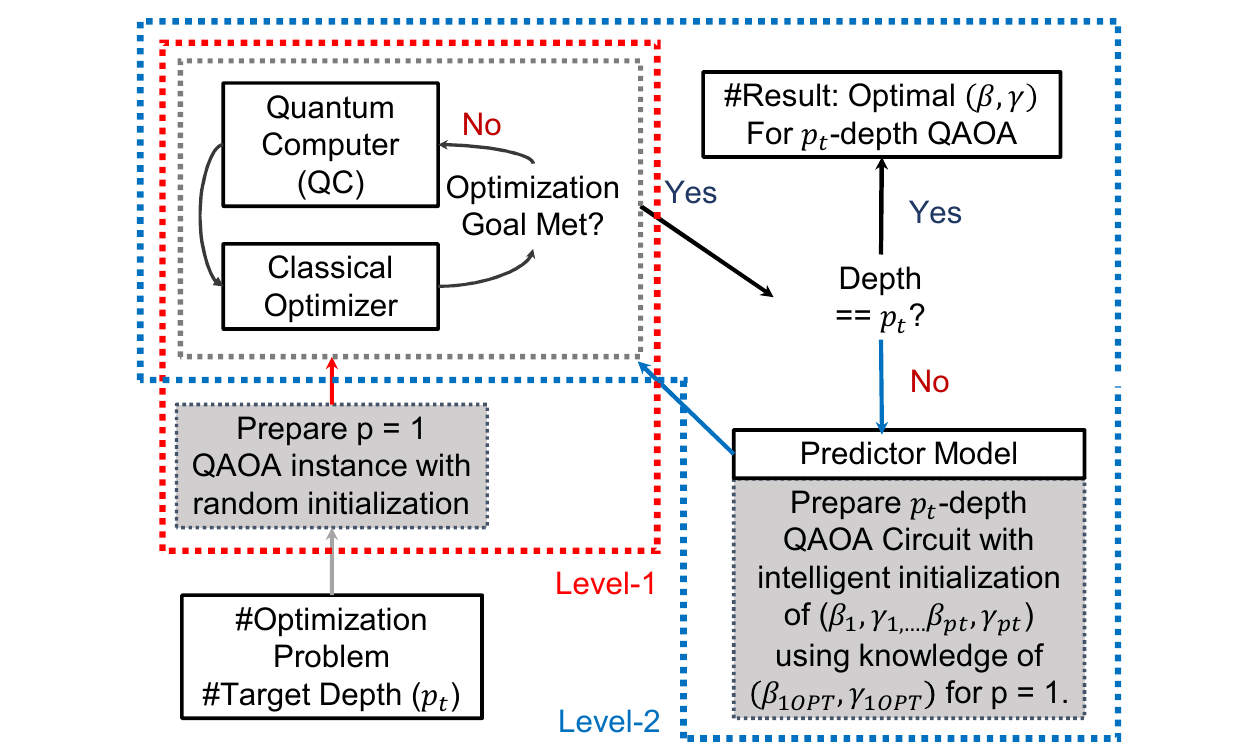}
 \end{center}
 \vspace{-1em}
 \caption{Proposed two-level QAOA implementation flow.} \label{fig:flow}
\end{figure}

To accelerate the convergence speed of QAOA with a classical local optimizer at moderate target depth-$p_t$, we propose a two-stage optimization procedure (Fig. \ref{fig:flow}). In the first stage, the QAOA instance of depth $p$ = 1 is optimized for the target problem. This is relatively simple and fast optimization process. The optimal $\gamma_{1OPT}(p=1)$ and $\beta_{1OPT}(p=1)$ values and the target depth-$p_t$ is then used to predict the initial values of the control parameters for the $p_t$-depth QAOA instance using pre-trained regression models. The local optimizer then optimizes the control variables from these initial values to find the target solution. The overall convergence speed is the summation of the number of loop iterations for $p$ = 1 instance optimization (typically fast) and the later target depth-$p_t$ instance optimization (which we have accelerated).

\section{Machine Learning Framework}
We train regression models with the optimal parameter values of various problem instances. 
To evaluate the proposed technique, we have created a data-set which includes the optimal QAOA parameter values of an ensemble of problem graphs for varied depths. The details are provided below.

\subsection{Data Generation}
We have picked the problem graphs for MaxCut-QAOA from the Erdos-Renyi ensemble \cite{bollobas2001random} with edge probabilities of 0.5 using Python NetworkX package \cite{developers2010networkx}. The ensemble is frequently used in graph theory to validate if a certain property holds for almost all types of graphs \cite{bollobas2001random}. A total of 330 graphs are chosen each containing 8 nodes with random number of edges and connectivity. The optimal control parameters for each of the graphs have been generated for various circuit depths ($p$ = 1 to 6) using L-BFGS-B as the classical optimizer \cite{jones2016scipy}. The quantum computer is simulated using the QuTIP framework \cite{johansson2013qutip}. The functional tolerance limit is identical for all the runs $(10^{-6})$ with optimization domain restricted to $\beta_i$ $\in$ [0,$\pi$], $\gamma_i$ $\in$ [0,2$\pi$] \cite{farhi2014quantum} for random initializations. The data is used to create the data-set (discussed next). \emph{Note that the data generation is an one-time cost.} 

\subsection{Data-set Analysis} \label{DATA}

\begin{figure} [] 
\vspace{-1em}
 \begin{center}
    \includegraphics[width=0.45\textwidth]{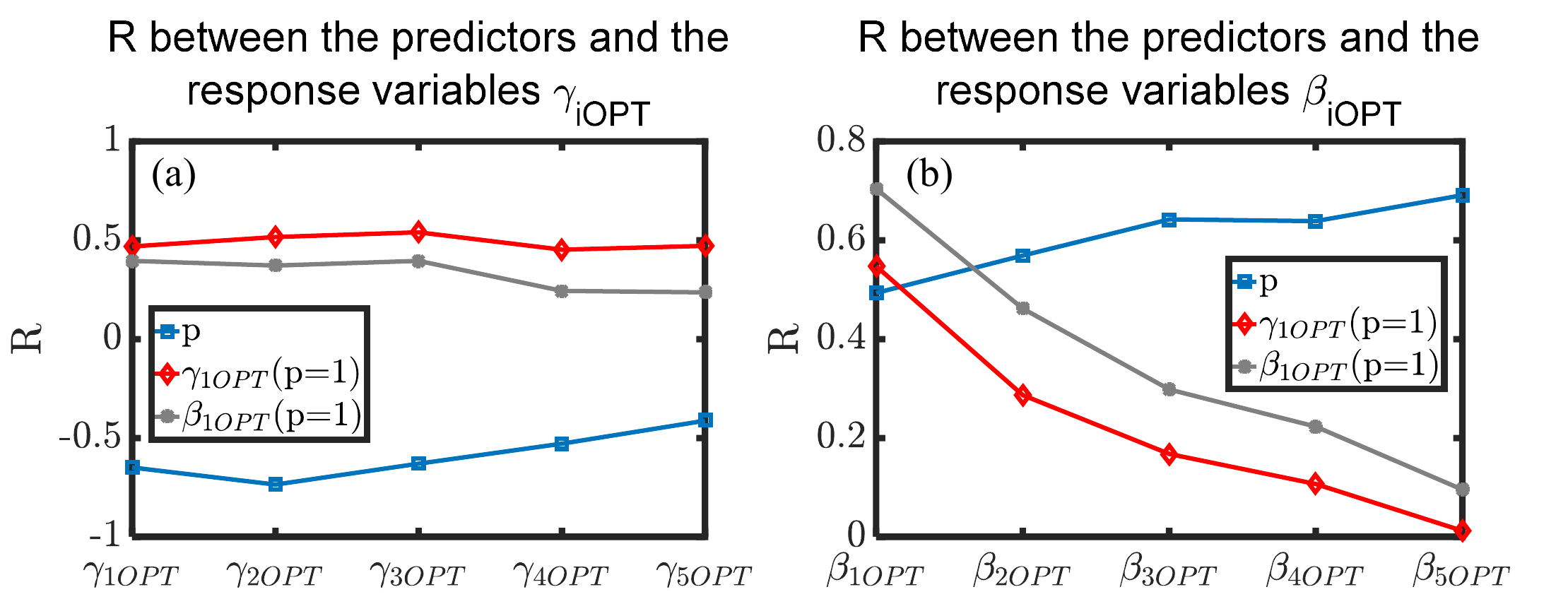}
 \end{center}
 \vspace{-1em}
 \caption{Correlation between the predictors of the two-level approach ($\gamma_{1OPT}(p=1)$, $\beta_{1OPT}(p=1)$, $p$) and the response variables for (a) $\gamma_{iOPT}$; (b) $\beta_{iOPT}$.} \label{fig:Corr}
 \vspace{-0.5em}
\end{figure}

We performed a detailed analysis of the correlations between the predictor (i.e., the input to the predictor model) and the response (i.e., the output of the predictor) variables in our data-set. The optimal $\gamma_{1OPT}$ and $\beta_{1OPT}$ parameters for $p$ = $1$ show a strong linear correlation (correlation coefficient, $R$ = 0.92) \cite{hastie2005elements} with each other. The correlation coefficient (R) between the $\gamma_{iOPT}$ and $p$ is negative which indicates a decrease in $\gamma_{iOPT}$ with increase in $p$ (Fig. \ref{fig:Corr}(a)). Note that the correlation decreased for higher order parameters. For instance, R between $\gamma_{1OPT}$ and $p$ is found to be -0.63, while it decreased to -0.44 for $\gamma_{5OPT}$. The correlation between $\beta_{iOPT}$ and $p$ is found to be positive and it increased for higher order parameters (Fig. \ref{fig:Corr}(b)). On one hand, the higher order $\beta_{iOPT}$ parameters (i.e. $\beta_{5OPT}$) are smaller compared to the lower order ones. On the other hand, the lower order $\gamma_{iOPT}$ parameters are smaller compared to the higher order ones. The results indicate that the $\gamma_{iOPT}$ and $\beta_{iOPT}$ values are more dictated by the $p$ values when those optimal parameters are expected to be small.

A decreasing trend is also observed in the R between $\beta_{iOPT}$'s and $\beta_{1OPT}(p=1)$, $\beta_{iOPT}$'s and $\gamma_{1OPT}(p=1)$, $\gamma_{iOPT}$'s and $\gamma_{1OPT}(p=1)$, and $\gamma_{iOPT}$'s and $\beta_{1OPT}(p=1)$ (Fig. \ref{fig:Corr}(a) and (b)). These variables showed positive correlations. The decreasing trend in these correlation coefficient indicates that the further we move from a certain depth, the associated control parameters will be weakly correlated. In other words, the control parameters at $p$ = 1 will be weakly correlated with the control parameters at $p$ = 3, compared to the parameters at $p$ = 2. We can expect high correlation between the optimal parameters at closer depths. 

The data-set is split into a training and a test data-set in 20:80 ratio (66 in the training data-set and 264 in the test data-set). The reason behind choosing a small training data-set is twofold - first, to determine whether the proposed approach generalizes to other instances beyond the training set; second, to emphasize that a small training-set is sufficient to reap the benefit of our proposed methodology.

\subsection{Supervised ML Models}

We have experimented with four different regression models - GPR, LM, RSVM and RTREE from MATLAB Statistics and ML Toolbox \cite{mathworks2017matlab} to analyze their performance as our predictor models. GPR showed the best performance metrics (lowest mean square error (MSE), root mean square error (RMSE), mean absolute error (MAE), and highest $R^2$ and $R^2_{adj}$ statistics \cite{hastie2005elements}) for all the tasks. Therefore, we have used GPR as our regression model in all further analysis. Note that, we have used MSE as the cost function during the training phase. 
The performance improvement with our proposed approach are summarized in the following Section.

\section{Numerical Simulation Results}

As a test-case, we have selected QAOA circuits with target depths ranging from $2$ to $5$ (i.e., a total of $4$, $6$, $8$, and $10$ parameters, respectively in the circuit). 

\begin{figure} [hb] 
 \begin{center}
    \includegraphics[width=0.49\textwidth]{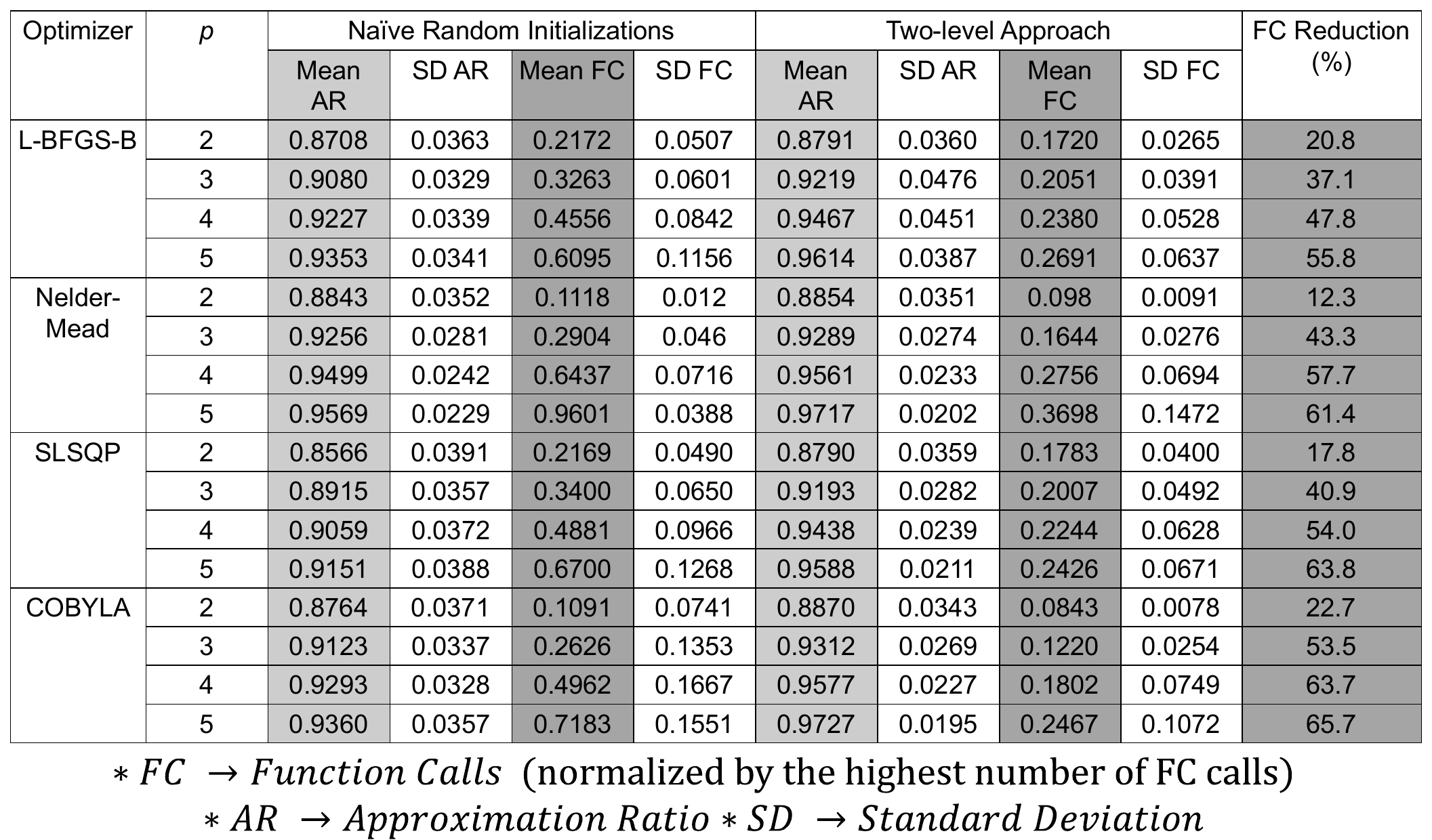}
 \end{center}
 \vspace{-1em}
 \captionof{table}{Run-time comparison between the random approach (Naive) and the two-level approach for L-BFGS-B, Nelder-Mead, SLSQP and COBYLA optimizers.} \label{fig:two_1}
 \vspace{-0.5em}
\end{figure}

First, we calculate the run-time for random initialization. For that, we solve the MaxCut problem for $264$ graphs each with following setup: $4$ different local optimizers, $20$ random loop initializations, circuit depth $p = 2$ to $5$, and tolerance $10^{-6}$. We report the mean and standard deviations of function calls (FC) (i.e., the number of loop iterations) along with the mean and standard deviations of approximation ratio (AR) in Table \ref{fig:two_1} for each depth and the local optimizer. 

Next, we calculate the run-time for our ML-based initialization. We solve the same MaxCut problem for $264$ graphs with identical setup (i.e., same local optimizers, circuit depth, and tolerance) except, the optimization loop is initialized with predicted parameters by the trained ML model. We report function calls and approximation ratios as before. The number of function calls in this case consists of two components:  \emph{number of calls to get $\gamma_{1OPT}(p=1)$ and $\beta_{1OPT}(p=1)$ (with random initialization) + number of calls to solve a problem for the target depth (with ML-based initialization).}

\begin{figure} [] 
 \begin{center}
    \includegraphics[width=0.48\textwidth]{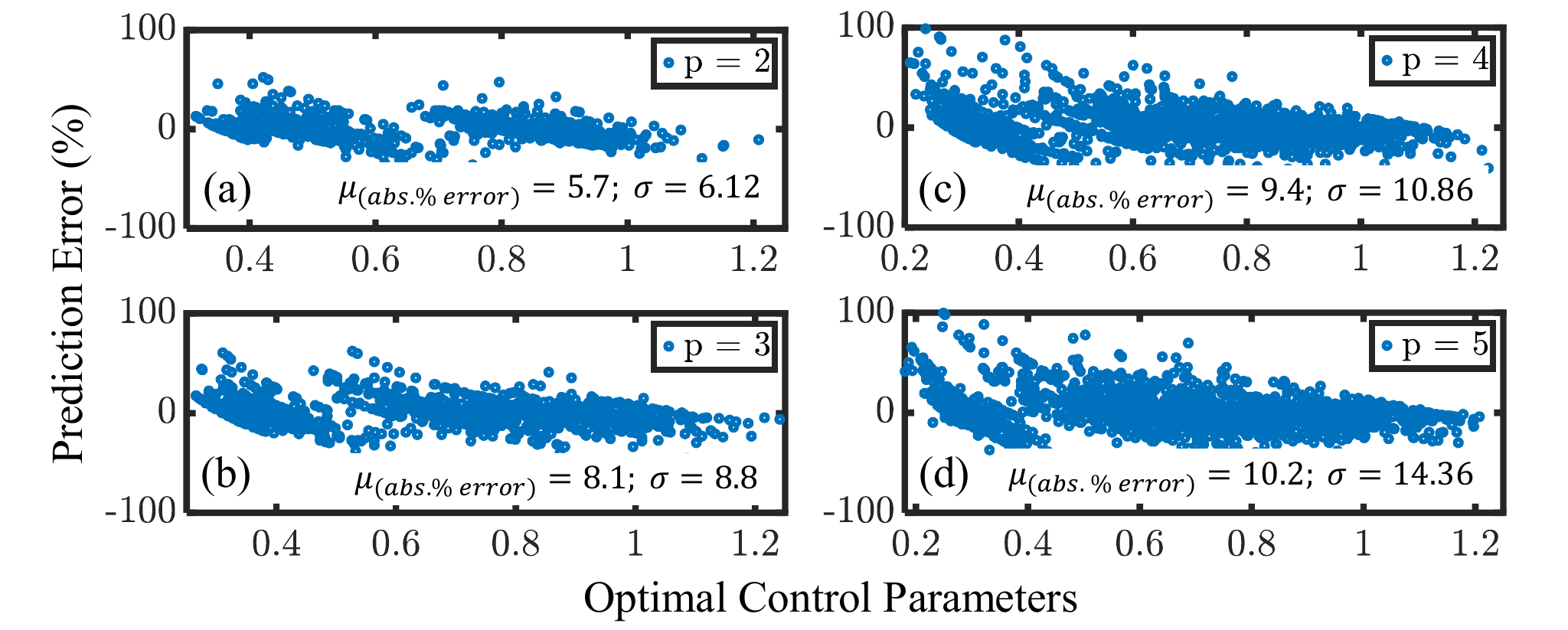}
 \end{center}
 \caption{Prediction errors in (a) $p$ = 2; (b) $p$ = 3; (c) $p$ = 4; and (d) $p$ = 5 QAOA instance control parameter predictions for the test data-set (264 graphs) in the two-level approach.} \label{fig:reg_inaccu}
 \vspace{-2em}
\end{figure}

Note that the prediction error is higher for larger target depths for the test data-set (Fig. \ref{fig:reg_inaccu}) e.g., the average percentage error in $p$ = 2 instance parameter predictions (difference from the actual optimal control parameter values for the 264 graphs in the test data-set) has been found to be 5.7\%. This is smaller than the errors in $p$ = 3 instance parameter predictions (8.1\%). The results match with our expectation as the predictor variables have weaker correlations with higher order control parameters (refer to Section \ref{DATA}).

The two-level approach shows an average improvement of 44.9\% in run-time across all the local optimization procedures. Note that the improvement in run-time is more pronounced when we go for a higher target depth implementation. For instance, Nelder-Mead optimizer showed an average reduction of 12.3\% in function calls over the naive approach for target depth $p$ = 2 for the entire test set and it increased to 43.3\% for $p$ = 3, and further increased to 57.7\% for $p$ = 5 (Table \ref{fig:two_1}). Note that, for any target depth-$p_t$, $p$ = 1 instance optimization constitutes a large portion of the total run-time (as it starts from random initialization). Therefore, the improvement in the runtime is less evident for the lower target depth (say, $p_t$ = 2) even though the prediction accuracy is higher.

\section{Conclusion}
In this paper, we presented an ML-based method to accelerate QAOA optimization loop using MaxCut problem. We present analysis for a broad-set of graphs, local optimizers, and circuit depths. We extract trends in the QAOA circuit parameters, select appropriate features for ML models, generate training data-set, and run QAOA optimization loop with predicted parameters. Our approach reduces the number of loop iteration by $44.9\%$ in average thus accelerating the process. We also present possible tweaks to augment our approach, relevant discussions and limitations, and future perspectives.

\noindent \textbf{Acknowledgement:} This work is supported by SRC (2847.001), and NSF (CNS- 1722557, CCF-1718474, CNS-1814710, DGE-1723687 and DGE-1821766).
\bibliographystyle{IEEEtran}
\bibliography{IEEEabrv,biblio}

\end{document}